\documentclass[draft,numberedheadings]{aipproc}

\layoutstyle{6x9}

\begin{document}

\def\simg{\mathrel{\rlap{\raise 0.511ex \hbox{$>$}}{\lower 0.511ex \hbox{$\sim$}}}}
\def\siml{\mathrel{ \rlap{\raise 0.511ex \hbox{$<$}}{\lower 0.511ex \hbox{$\sim$}}}}
\def\epsi{\epsilon_e} \def\epsB{\epsilon_B} \def\cm3{\,\rm cm^{-3}}
\def\etal{et al$.$ } \def\eg{e.g$.$ } \def\ie{i.e$.$ } \def\deg{^{\rm o}} \def\E{{\cal{E}}}

\title{ Phases of Swift X-ray Afterglows }

\author{A. Panaitescu}{ address={ ISR-1, Los Alamos National Laboratory, Los Alamos, NM 87545, USA} }

\begin{abstract}
 The X-ray afterglows observed by Swift exhibit rich light-curves, with four phases of different decay 
rate. The temporal and spectral properties for a set of 47 bursts
are used to identify the mechanisms which can explain these four phases.
The early, fast-decaying phase can be attributed to the same mechanism which generated the burst emission
(internal shocks in a relativistic outflow), while the following phases of slower decay can be identified
with synchrotron emission from the forward shock sweeping the circumburst medium. Most likely, the phase 
of slowest decay is due to a continuous energy injection in the forward shock. That the optical power-law
decay continues unabated after the end of energy injection requires an ambient medium with a wind-like 
density structure ($n\propto r^{-2}$) and forward shock microphysical parameters that change with the 
shock's Lorentz factor. A later break of the X-ray light-curve can be attributed to a collimated outflow 
whose boundary becomes visible to the observer (a jet) but the optical and X-ray decays are not always 
consistent with the standard jet model expectations.
\end{abstract}

\maketitle

\begin{footnotesize}

\section{Introduction}
\label{in}

\hspace*{3mm} 
 It is widely believed that the highly variable emission of a GRB arises in an ultrarelativistic outflow
(to explain its optical thinness to pair formation \& electron scattering and yield a non-thermal burst
spectrum extending over 1 MeV, sometimes even 1 GeV) where dissipation of the kinetic energy (and subsequent 
particle acceleration \& emission of synchrotron/inverse Compton radiation) is caused by large fluctuations
in the Lorentz factor (the internal shock model -- Rees \& M\'esz\'aros 1994). The X-ray, optical and radio 
emission following the burst is attributed to the interaction between the GRB ejecta and the circumburst 
medium, which leads to a short-lived reverse shock crossing the GRB ejecta and a much longer-lived 
forward shock energizing the ambient medium (the external shock model - \eg Paczy\'nski \& Rhoads 1993,
M\'esz\'aros \& Rees 1997). 

 In its simplest form, the standard external (forward) shock model assumes a GRB outflow with a constant 
shock energy, a uniform kinetic energy per solid angle, and constant microphysical parameters. The possibility
of energy injection into the forward shock was proposed by Paczy\'nski (1998) and Rees \& M\'esz\'aros (1998), 
while the effect of an angular structure of the outflow (where the energy per solid angle is not constant) was 
investigated for the first time by M\'esz\'aros, Rees \& Wijers (1998). 

 The first possible evidence for energy injection in the forward shock was provided by the rise of the optical 
emission of the GRB afterglow 970508 after 1 d (Panaitescu, M\'esz\'aros \& Rees 1998). Further evidence for 
this process may be the slow optical decay of the afterglows 010222 at 0.1--1 d (Bjornsson \etal 2002) and
021004 before 0.1 d (Fox \etal 2002) and the fluctuations in the optical emission of GRB afterglow 030329 after 
1 d (Granot, Nakar \& Piran 2003). 

 The effect of GRB ejecta collimation was treated in detail by Rhoads (1999), and shortly after that the 
optical light-curve of the GRB afterglow 990123 exhibited the predicted break (Kulkarni \etal 1999). 
Since then, about a dozen of other optical afterglows displayed a break at around 1 d (\eg Zeh, Klose 
\& Kann 2006), which has been interpreted as evidence for GRB jets. 
As pointed out by Rossi, Lazzati \& Rees (2002) and by Zhang \& M\'esz\'aros (2002) such light-curve breaks 
may also arise from structured outflows seen off-axis. 

 Cessation of energy injection cannot explain the optical light-curve breaks and the radio \& X-ray emission 
properties at the same time, but emission from the reverse shock crossing the incoming ejecta may accommodate 
the slow radio decays seen in a few cases after about 10 d (Panaitescu 2005). 

 Prior to Swift, broadband afterglow observations could be explained by the forward-shock model with 
constant microphysical parameters. As discussed in \S\ref{evol}, if optical and X-ray emission arise from 
the same outflow then Swift observations of a chromatic X-ray light-curve break at about 1 h after trigger
require that this basic assumption of the standard forward-shock model is abandoned.

 The results presented in the following sections are based on a sample of 37 afterglows observed mostly
by Swift until October 2005. The sample is nearly the same as that presented by O'Brien \etal (2006).

\section{GRB 050315 - a canonical X-ray afterglow}

\begin{figure}
    \parbox[h]{9cm}{ \includegraphics[height=7cm,width=9cm]{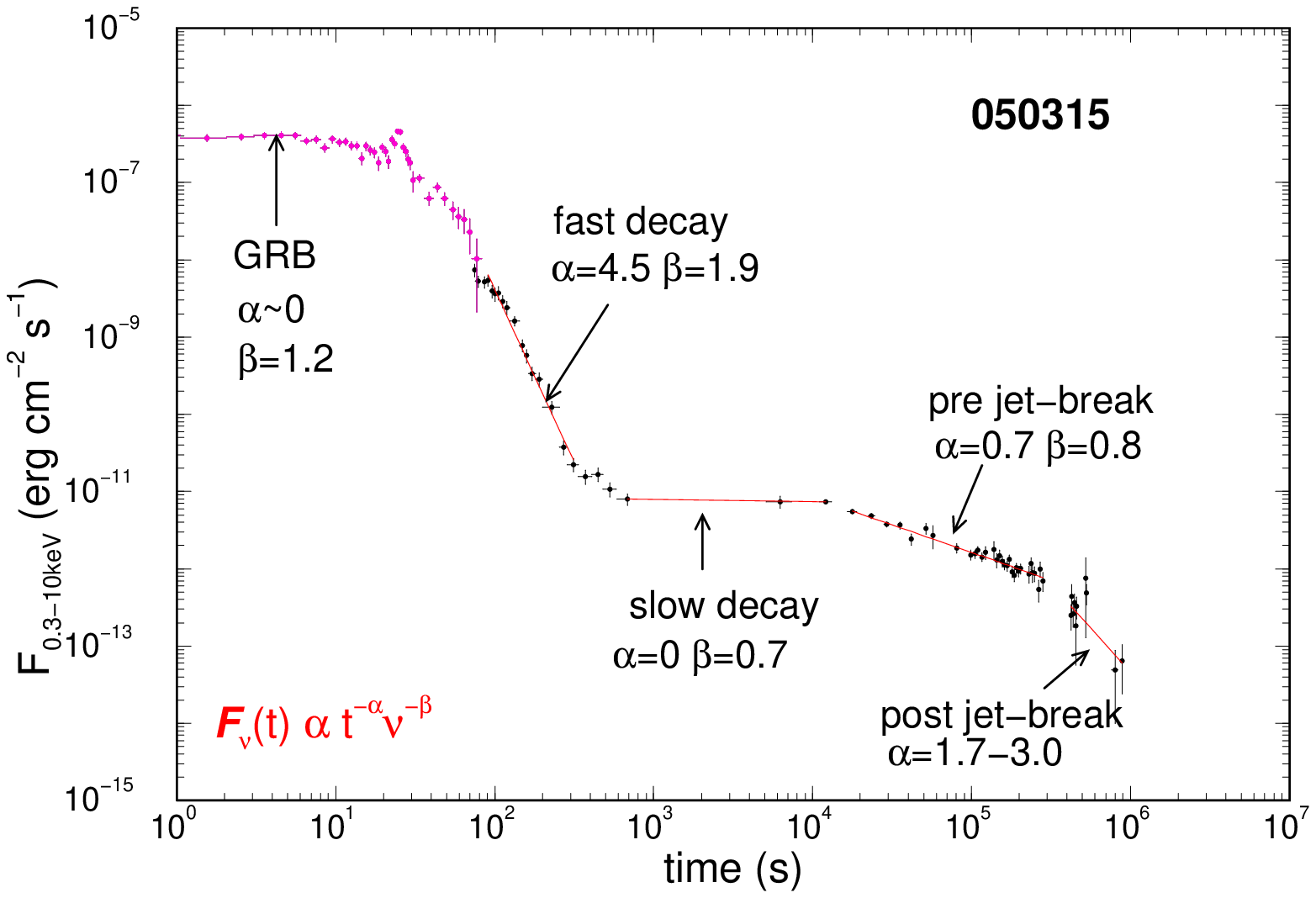} }
    \hspace*{2mm}  \parbox[h]{5.5cm}{ \footnotesize
        Phases of GRB afterglow 050315 and power-law fits to each decay phase. Red points indicate
        15--150 keV BAT measurements extrapolated to the 0.3--10 keV XRT band assuming that the burst 
        power-law spectrum extends unbroken below 15 keV. Black points indicate XRT measurements. }
\caption{}
\label{0315}
\end{figure}

\hspace*{3mm} 
 GRB 050315, shown in figure \ref{0315}, is a typical burst followed by an X-ray afterglow which displays 
all the characteristic phases of Swift afterglows with the exception of flares. The 15--150 keV emission 
of the long bursts observed by Swift lasts 10--100 s, has a spectral energy distribution $F_\nu \propto 
\nu^{-\beta}$ of average slope $\overline{\beta_\gamma} = 0.7\pm0.5$. 
In chronological order, the phases of the ensuing afterglow are :
\begin{enumerate}
 \item the {\sl fast-decay phase}, was observed in two pre-Swift GRBs (990510 and 010222), but has not been 
   clearly identified because the following slow-decay phase was not monitored. 
   It extends over a factor 3--30 in time and, during it, the 0.3--10 keV flux falls-off by 1--4 orders of 
   magnitude. There are roughly equal numbers of afterglows for which this phase starts in one of the following 
   time intervals: $t < 10$ s, $t=10-30$ s, $t=30-100$ s, and $t > 100$ s. The average spectral slope of the 
   0.3--10 keV emission during this phase is $\overline{\beta_{x1}} = 1.0 \pm 0.5$. It can be identified with 
   the burst's internal-shock emission,
 \item the {\sl slow-decay phase}, was observed for the first time by Swift, and lasts over a factor 10--100 
   in time, until $10^3 - 5\times 10^4$ s.
   During this phase, the average spectral slope is $\overline{\beta_{x2}} = 0.9 \pm 0.2$ and the emission 
   power-law decay, $F_x \propto t^{-\alpha_x}$, has an average index $\overline{\alpha_{x2}} = 0.6 \pm 0.3$,
   being smaller (\ie slower decay) than expected in the standard forward-shock model. 
   For about 30\% of afterglows, the fast-decay and slow-decay phases are not manifested, instead one
   single phase of moderate decay rate is observed, 
 \item the {\sl pre jet-break phase}, ending sometimes during the first day, but lasting for tens of days
   in other cases, has an average decay index $\overline{\alpha_{x3}} = 1.25 \pm 0.25$. The X-ray light-curve 
   decay is consistent with the prediction of the standard forward-shock model if the angular extent of 
   the visible part of the outflow is less than the true outflow opening, \ie if the observer does not see 
   yet the ejecta boundary in the direction perpendicular to their motion,
 \item the {\sl post jet-break phase}.
   XRT observations do not extend usually late enough to cover this phase of fast decay.
   A few afterglows display it at a few hours to days after trigger but there are also a couple of X-ray 
   afterglows for which the pre jet-break phase extends up to at least 10 days. 
\end{enumerate}
The afterglow emission of the last three phases arises from the external forward-shock, but the slow-decay
phase points to a departure from the standard model.

 About half of the well-monitored X-ray afterglows exhibit flares, predominantly in the first hour after 
the burst (Burrows \etal 2005). Of those 19 flaring X-ray afterglows, two-thirds exhibit flares during the 
fast-decay phase and nearly half of them show flares during the slow-decay phase. It is natural to attribute
the flares during the fast-decay phase to the same internal shocks. There are three facts which indicate 
that flares occurring later, during the slow-decay phase, also originate from internal shocks and not
from the same mechanism as the underlying X-ray afterglow at that time: 
\begin{enumerate} 
 \item The optical emission of GRB afterglow 050904 does not track the $10^4-4\times 10^4$ s 
       X-ray flare(s) (fig. 1 of Gendre \etal 2006), the same being true for the 100--500 s X-ray flares
       of GRB afterglow 060210 (fig. 4 of Stanek \etal 2006),
 \item The flare of GRB 050502B has a spectrum harder than that of the underlying emission (fig. 4 of 
       Falcone \etal 2006),
 \item The duration of flares ($\delta t /t \siml 1$) is shorter than the spread in the photon arrival-time
       over the opening of the outflow which is visible to the observer and are, sometimes, too sharp to
       be explained by an inhomogeneity in the circumburst medium with an angular scale smaller than the
       visible outflow opening (fig. 4 of Zhang \etal 2006, figs. 13 \& 14 of Nakar \& Granot 2006).
\end{enumerate}

\section{GRB emission}

\hspace*{3mm} 
 Perhaps the greatest difficulty that the internal-shock 
model encounters is the large efficiency at which the outflow kinetic energy seems to be converted into 
$\gamma$-rays. This requires large fluctuations in the ejecta Lorentz factor but it is not sufficient to
explain the high (usually above 25\%) overall GRB efficiency, as random fluctuations in the outflow Lorentz 
factor channel only a small fraction ($\siml 10$\%) of the radiated emission into the 15--350 keV window.

 For the typical GRB pulse duration, $\delta t\sim  1$ s, the internal shocks should occur at a distance 
$\Gamma^2 c \delta t \sim 1.2\times 10^{15} (\Gamma/200)^2$ cm, where $\Gamma$ is the ejecta Lorentz factor.  
The radius at which the GRB outflow is decelerated by the interaction with the Wolf-Rayet (the progenitor
of long bursts) wind is 
$1.0\times 10^{15} (\Gamma/200)^{-2}$ cm, for an isotropic-equivalent outflow energy of $10^{53}$ ergs.
This shows that outflow Lorentz factors above 200 would lead to a deceleration of the ejecta occurring 
before internal shocks, \ie most collisions would be with the decelerating leading edge of the outflow. 
This internal--external shock model (Fenimore \& Ramirez-Ruiz 1999) has a higher dissipation efficiency 
than that of the internal-shocks model but may lead to an increase of the pulse duration with time which
is in contradiction with the observed constancy of the pulse duration during the burst (Ramirez-Ruiz \&
Fenimore 1999), which suggests that the outflow Lorentz factor does not greatly exceed 200.

 If the same Lorentz factor where below 50, then the observer-frame timescale for the outflow deceleration, 
$300 (\Gamma/50)^{-4}$ s, would be longer than the latest time when the forward-shock emission is seen to 
emerge, at about 400 s. That this emerging emission is not rising implies that the deceleration timescale 
is less than 300 s and that the ejecta Lorentz factor is larger than 50. (Furthermore, Lorentz factors 
much below 100 would may make the ejecta optically thick to electron scattering at the radius of internal
shocks, which would either yield a thermal burst spectrum, if there are cold electrons in the ejecta, or 
a non-thermal burst spectrum peaking above 1 GeV, if all electrons are relativistic).

 Therefore, the average wind Lorentz factor should be around 100, within a factor of 2--3, and the angular 
extent of the outflow region from which the observer receives the GRB emission is $\Gamma^{-1} \sim 0.01$ 
rad, \ie about $0.5\deg$.

\section{Fast-decay phase}

\hspace*{3mm} 
 In all bursts with a continuous monitoring of the fast-decay phase, the extrapolation of the 15--150 keV BAT 
light-curve to the 0.3--10 keV XRT band matches the XRT light-curve at the time when observations
by both instruments overlap. When a temporal gap exists between observations made with the two instruments, 
a power-law temporal extrapolation of the BAT GRB tail almost always matches the XRT flux at the beginning
of the X-ray afterglow, or the back extrapolation of the XRT early afterglow flux matches the BAT flux at the 
and of the GRB. This indicates that the burst and fast-decay afterglow emission arise from the same process.

\vspace*{5mm}

 One possible reason for this fast decay is that the burst emission ceased so abruptly that what we see during 
this phase is the GRB emission from the shocked gas moving at angles $\theta$ larger than the $\Gamma^{-1}$ 
visible during the burst phase (Kumar \& Panaitescu 2000). Then there is a one-to-one correspondence between 
the photon arrival-time $t$ and $\theta$; the relativistic boost ${\cal D}$ of the photon energy and $t$ satisfy
\begin{equation}
  t \propto \theta^2 \;,\; {\cal D} \propto \theta^{-2} \;.
\label{theta}
\end{equation}
Thus, the emission arriving at later times is less beamed relativistically, less boosted in energy, and less 
compressed in observer time. If the burst emission is uniform within an angle of ${\rm few}/\Gamma$, then 
the large-angle GRB emission exhibits a power-law decay of index 
\begin{equation}
   \alpha_{x1} = 2 + \beta_{x1} \;,
\label{a1}
\end{equation}
where $\beta_{x1}$ is the afterglow X-ray spectral slope.

 Figure \ref{x1} shows the X-ray decay index and spectral slope of 28 afterglows whose fast-decay phase 
was well-monitored, in the sense that the gap between the last BAT and first XRT measurements is less than 
a factor 10. Bursts with a wider gap or whose very early X-ray emission is dominated by flares were excluded. 
For the bursts shown in figure \ref{x1}, the linear correlation coefficient of $\alpha_{x1}$ and 
$\beta_{x1}$ is $0.63\pm0.10$, corresponding to less than 0.4\% probability of a chance correlation. 
For more than half of bursts the fast-decay  satisfies equation (\ref{a1}). For about a quarter, the early 
X-ray emission falls-off faster than expected for the large-angle GRB emission, and a few afterglows exhibit 
a slower decay.

\begin{figure}
    \parbox[h]{8.5cm}{ \includegraphics[height=7cm,width=8.5cm]{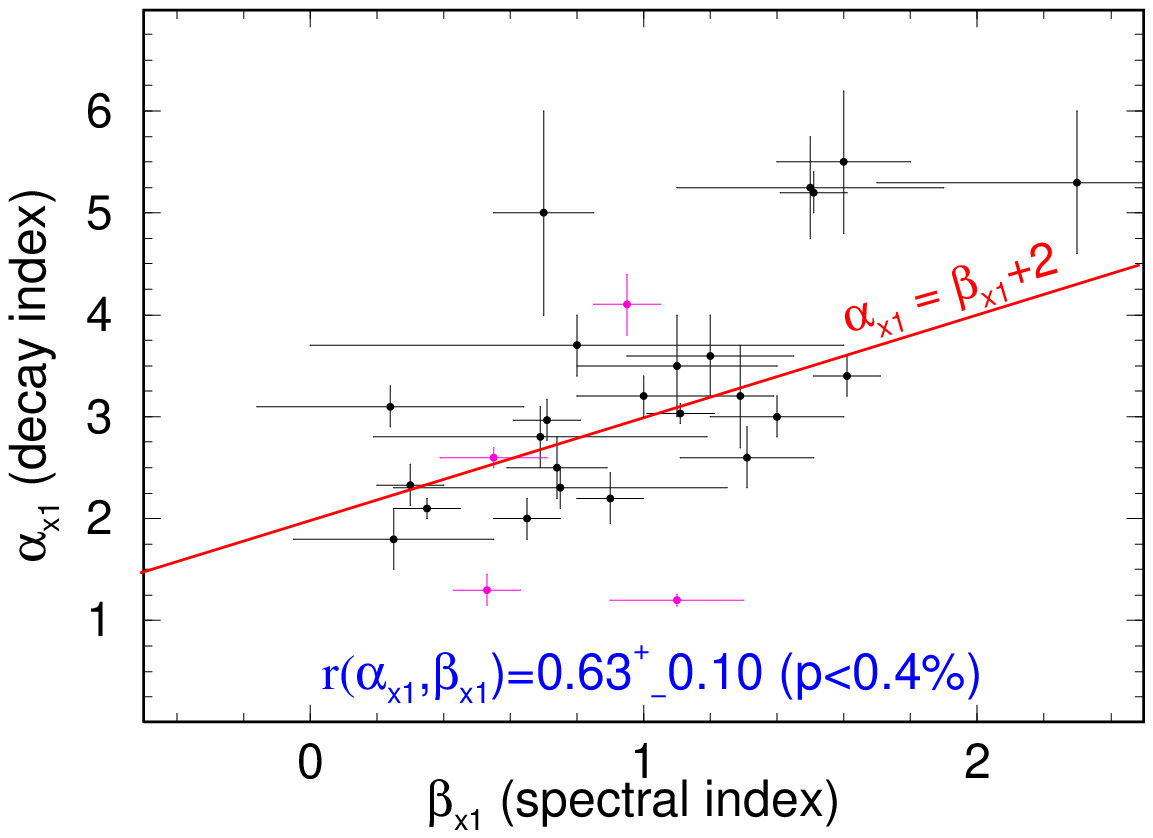} }
    \hspace*{2mm}  \parbox[h]{6cm}{ \footnotesize
        Comparison of the decay index \& spectral slope during the fast-decay phase and the prediction 
        of the large-angle GRB emission model (red line). Black symbols for long bursts, red symbols
        for short bursts. }
\caption{}
\label{x1}
\end{figure}

 Fig. 3 of Zhang \etal (2006) shows that the decay of the large-angle emission appears steeper if the trigger
time underestimates the beginning time for the last GRB pulse (which dominates the early GRB tail emission),
accounting thus for the afterglows with a faster-than-expected decay. After twice the burst duration, this 
effect should be negligible and the decay given in equation (\ref{a1}) should set-in. This is in fact observed:
all the steep decays shown at the top of figure \ref{x1} are seen only over a two-fold increase in time. 
Alternatively, a steeper-than-expected decay of the large-angle emission can be due to the luminosity of the 
emitting surface decreasing with angle at $\theta > \Gamma^{-1}$, \ie to a GRB outflow having a structure on 
an angular scale of $\sim 0.5\deg$. 

 For 40\% of bursts, the 0.3--10 keV spectral slope during the fast-decay phase is comparable to that during
the burst while for about half of afterglows the early X-ray emission is softer than that of the burst
(figure \ref{bgbx1}). The average ${\beta_{x1}} -\beta_\gamma$ is 0.3 with a dispersion of 0.5 .  
One possible explanation is that the last pulse of the burst, which dominates the fast-decay phase emission 
from the end of the burst ($t_\gamma$) to about $2t_\gamma$, is softer than the GRB, which is consistent with 
the well-known softening trend of the burst emission (\eg Norris \etal 1986). Another possibility is that
the GRB spectrum becomes softer over an angular scale of order $\Gamma^{-1}$, but that would require that 
only bursts whose patch of hardest emission is moving directly toward the observer are detected. 
Still another possibility is that the GRB emission has no angular structure but the spectrum of internal-shocks 
has a break at about 10 keV, above which the emission is harder (right panel of figure \ref{bgbx1}). 
The relativistic boost of photon energy being inversely proportional to the photon arrival-time (equation 
\ref{theta}) means that during the GRB phase ($t \siml t_\gamma$), the 15--150 keV BAT band corresponds 
to $(15-150)/\Gamma$ keV in the comoving frame of the emitting gas, while during the fast-decay phase 
(from $t=t_\gamma$ to $t \sim 3 t_\gamma$), the 0.3--10 keV XRT band probes the range $(0.3-10)/\Gamma$ keV 
to $(1-30)/\Gamma$ keV. This explanation cannot be at work for fast-decay phases extending over more than
$3 t_\gamma$ as it would imply a spectral break above 10 keV, which would be at odds with BAT observation 
of burst SEDs.

\begin{figure}
  \includegraphics[width=15cm,height=6.5cm]{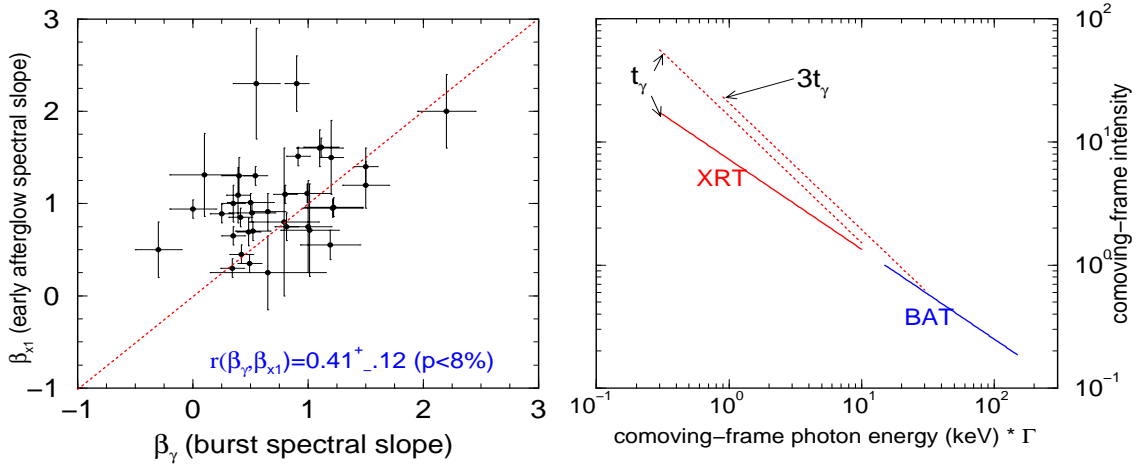} 
\caption{ Left panel: Comparison of the spectral slopes during the burst and afterglow fast-decay phase.
     For half of afterglows, the burst emission is harder than during the fast-decay phase.
   Right panel: photon energy ranges which are relativistically boosted in the BAT and XRT bands.
     Dotted lines indicate an afterglow spectrum softer than that of the burst. }
\label{bgbx1}
\end{figure}

\vspace*{5mm}

 An alternative explanation for the fast-decay phase is that it arises from bright spot on the outflow 
surface, of angular opening less than the $\Gamma^{-1}$ visible to the observer. This model has an advantage 
over models where the burst emission arises from a shell in that the absence of emission from angles of 
order $\Gamma^{-1}$ may can account for the time-symmetry of GRB pulses (Norris \etal 1996). Further,
the lack of the large-angle emission means that the fast X-ray decay reflects the off-switching of the spot 
emission after the cessation of internal shocks. Very steep decays could result after the cooling frequency
falls below the X-ray, due to the exponential cut-off of the synchrotron emissivity above its peak frequency.

\section{Slow-decay phase}
\label{slow}

\hspace*{3mm} 
 The definition for this phase is that it immediately follows the fast-decay phase or, if there is a 
temporal gap in afterglow monitoring, we require the existence of a subsequent faster decay which is slower 
than that of a spreading jet. The latter condition is imposed to avoid mis-identifying the slow-decay phase 
with the pre jet-break phase (\S\ref{pre}).   

 For a set of 14 afterglows with reported spectral slopes during this phase, $\beta_{x2} = 0.9 \pm 0.2$.
About one-third of them have $\beta_{x1} > \beta_{x2}$, while for the rest $\beta_{x1} = \beta_{x2}$ within 
the measurement errors. Thus there is evidence that the X-ray continuum hardens sometimes at the transition 
from the fast-decay to the slow-decay phases, which is not surprising if the emissions during these two phases 
arise from different mechanisms (internal-shocks and external forward-shock, respectively).

 Figure \ref{x2} shows the decay indices and spectral slopes for 26 X-ray afterglows during the slow-decay
phase, the 14 above plus 12 for which $\beta_{x2} = \beta_{x1}$ was assumed. Also shown is the $\alpha_{x2} -
\beta_{x2}$ dependence expected in the standard forward-shock model, for uniform and wind-like circumburst 
media and for the two possible locations of the cooling frequency ($\nu_c$) relative to the 0.3--10 keV XRT 
band. As can be seen, half of afterglows exhibit a decay which is slower than expected in the standard 
forward-shock model. For the other half, consistency between observations and model require that $\nu_c$ is 
below the X-ray and that the index of the electron distribution with energy in the forward shock is 
$p = 2\beta_{x2} < 2$ (dashed line in figure \ref{x2}). 

\begin{figure}
    \parbox[h]{8.5cm}{ \includegraphics[height=7cm,width=8.5cm]{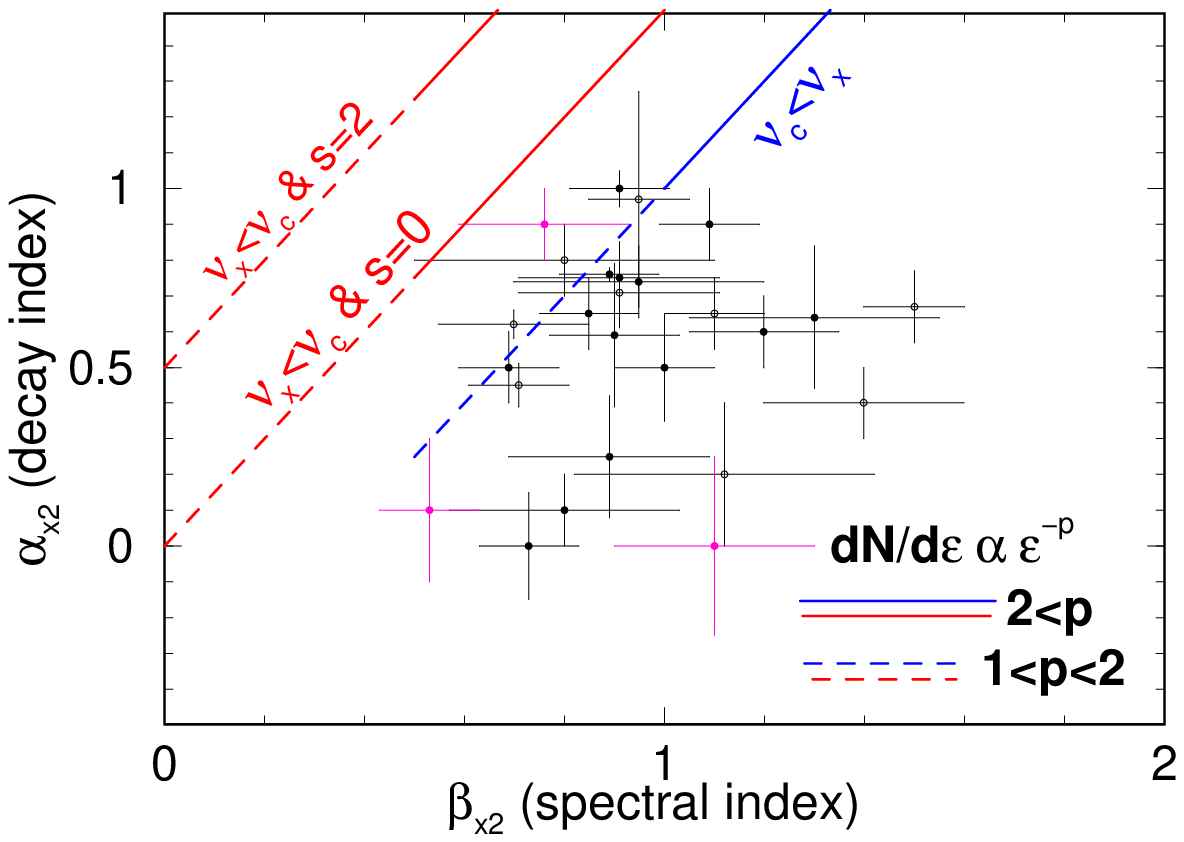} }
    \hspace*{2mm}  \parbox[h]{6cm}{ \footnotesize
        Comparison between decay index \& spectral slope measured during the slow-decay phase and the
        standard forward-shock model expectations, for a $n \propto r^{-s}$ stratification of the
        circumburst medium ($s=0$ for a homogeneous medium, $s=2$ for a wind). If the cooling frequency 
        is below X-ray ($\nu_c < \nu_x$) the decay index is independent of $s$. As can be seen, the
        decay during this phase is too slow for the standard forward-shock model. $p$ is the index
        of the electron distribution with energy in the forward shock. }
\caption{}
\label{x2}
\end{figure}

 Optical light-curves decay from the earliest observations (at $\sim 100$ s), indicating that the electron 
power-law distribution at that time must extend over at least $\log(10\,{\rm keV}/1\,{\rm eV})^{1/2} = 2$ 
decades in energy. In the standard forward-shock model (where microphysical parameters are constant), the 
energy of the electrons radiating at a fixed observing frequency (X-ray) increases as $t^{3/8}$ (for a 
homogeneous medium) or $t^{1/2}$ (for a wind-like medium). Given that the X-ray emission is detected for 
4 decades in time (until $10^6$ s or later), these scalings show that the power-law electron distribution 
must extend over an extra 2 decades in energy, \ie to a total of more than $n = 4$ dex. Numerical modelling 
of broadband afterglow emission shows that, for a dozen afterglows, the lowest electron energy corresponds 
to more than 1\% of the forward shock energy. Then, for $p < 2$, the total electron energy is a fraction 
$\simg 1\% \times 10^{(2-p)n}$ of the shock energy which, for some of the afterglows shown in figure \ref{x2} 
near the dashed blue line, implies a total electron energy above equipartition. For those afterglows there
should be a high-energy cut-off of the electron distribution corresponding to a sub-equipartition total 
electron energy that would lead to a break in the X-ray light-curve before $10^6$ s, accompanied by a spectral 
softening. The apparent lack of such spectral evolution in late X-ray afterglows suggests that, more likely, 
all the  X-ray decays shown in figure \ref{x2} are too slow for the standard forward-shock model.

 Therefore there must be a mechanism which mitigates the decay during this phase. Given the factors which 
determine the forward-shock synchrotron flux, that mechanism could be: 
\begin{enumerate}
 \item an increase of the forward shock's average kinetic energy per solid angle ($\overline{\E}$) over the 
       region ($\theta < \Gamma^{-1}$) visible to the observer.
       In this scenario, the increase of $\overline{\E}$ could be due to either injection of energy in the 
       forward shock or to a structured outflow where regions of higher $\E$ enter the ever-increasing part 
       of the outflow visible to the observer. 
 \item a growth of the microphysical parameters in the forward shock: the fractional energy of electrons and 
       of the magnetic field. Here, it is assumed that these parameters may depend on the shock's Lorentz factor.
\end{enumerate}

\subsection{Energy injection}
\label{enginj}

 A continuous energy of injection in the forward shock could result from 
\begin{enumerate}
  \item a short-lived engine which releases all the ejecta simultaneously at different Lorentz factors
        (ordered increasingly in the radial outward direction by internal shocks) so that they catch up 
        with the leading forward shock as it is decelerated by the interaction with the circumburst medium, or
  \item a long-lived engine which releases ejecta on a laboratory-frame timescale comparable to the
        observer-frame time when the energy injection occurs.
\end{enumerate}
 For a power-law distribution of the incoming-ejecta kinetic energy, $d\E/d\Gamma_i \propto \Gamma_i^{-(e+1)}$, 
it can be shown that the parameter $e$ which reconciles the observed X-ray decay index $\alpha_{x2}$ and 
spectral slope $\beta_{x2}$ is : 
\begin{equation}
 {homogeneous\;\; medium :}\quad
            e (\nu_x < \nu_c) = 4 \frac{3\beta_{x2}-2\alpha_{x2}}{\alpha_{x2}+3} \;,\quad
            e (\nu_c < \nu_x) = 4 \frac{3\beta_{x2}-2\alpha_{x2}-1}{\alpha_{x2}+2} 
\label{es0}
\end{equation}
\begin{equation}
 {wind\; medium :} \hspace*{12mm}
          e (\nu_x < \nu_c) = 2 \frac{3\beta_{x2}-2\alpha_{x2}+1}{\alpha_{x2}-\beta_{x2}} \;,\quad
          e (\nu_c < \nu_x) = 2 \frac{3\beta_{x2}-2\alpha_{x2}-1}{\alpha_{x2}-\beta_{x2}+1}  \;.
\label{es2}
\end{equation}

For the afterglows of figure \ref{x2} and a homogeneous medium, the resulting indices have an average 
$\bar{e} = 2.0 \pm 1.0$ for cooling frequency above X-ray and $\bar{e} = 1.3 \pm 1.4$ for cooling below 
X-ray. A wind-like medium leads to $e < 0$ for many afterglows, in which case the forward shock energy, 
$\E (> \Gamma_i) \propto \Gamma_i^{-e}$, increases only if the forward shock is accelerating. In this
case, energy injection would continue only if the Lorentz factor of the incoming ejecta increases, \ie
the ejecta Lorentz factor should be decreasing radially outward, which is in contradiction with what
internal collisions in such a wind would produce. For this reason, we consider in the following only a 
homogeneous medium.

  Because the X-ray decay after the slow-decay phase is generally consistent with the standard forward-shock 
model expectations (\S\ref{pre}), energy injection should be negligible after the end of the slow-decay 
(at $t_b$). This implies the existence of a cut-off $\Gamma_{break} \equiv \Gamma_i(t_b)$ in the Lorentz 
factor of the incoming ejecta which carry a dynamically-important kinetic energy. For a long-lived engine, 
$\Gamma_{break}$ can be arbitrarily higher than the Lorentz factor $\Gamma(t_b)$ of the forward shock, thus
$\Gamma_{break}$ is unconstrainable. The high GRB efficiency (\S\ref{effi}) favours a short-lived engine,
in which case it can be shown that the ratio of the Lorentz factor of the incoming ejecta to that of the 
forward shock is time-independent : 
\begin{equation}
 \frac{\Gamma_i}{\Gamma} = \sqrt{\frac{8+e-2s}{e+2}} = \frac{\Gamma_{break}}{\Gamma(t_b)} \;.
\label{Gi}
\end{equation}
This allows the determination of $\Gamma_{break}$ as $\Gamma(t_b)$ can be estimated from the blast-wave 
energy $\E$ and medium density $n$:
\begin{equation}
 \Gamma (t_b) \simeq 600 \left( \frac{\E_{51}}{n_0} \right)^{1/8} \left( \frac{t_b}{z+1} \right)^{-3/8}
\label{Gm}
\end{equation}
where $\E_{51}$ is the kinetic energy per solid angle measured in $10^{51}$ erg/sr, $n_0$ is the ambient
medium proton density in $\cm3$, and $z$ is the burst redshift. We assume that $n = 1 \cm3$ and that
the total ejecta energy injected until $t_b$ is the same as the burst output in the lab-frame 20--2000 keV  
band. The small power at which $\E$ appears in equation (\ref{Gm}) shows that the error introduced by
this approximation should be small, however a factor 10 uncertainty in the medium density translates into a
30\% error in $\Gamma_{break}$. 

 Figure \ref{ei} shows the distribution of the incoming ejecta cumulative energy for afterglows with known
redshift, obtained with the aid of equations (\ref{es0}), (\ref{Gi}), and (\ref{Gm}). The width of the
$\Gamma_{break}$-distribution is fairly large and there does not appear to be a universal value.

\begin{figure}
  \includegraphics[height=6.5cm,width=15cm]{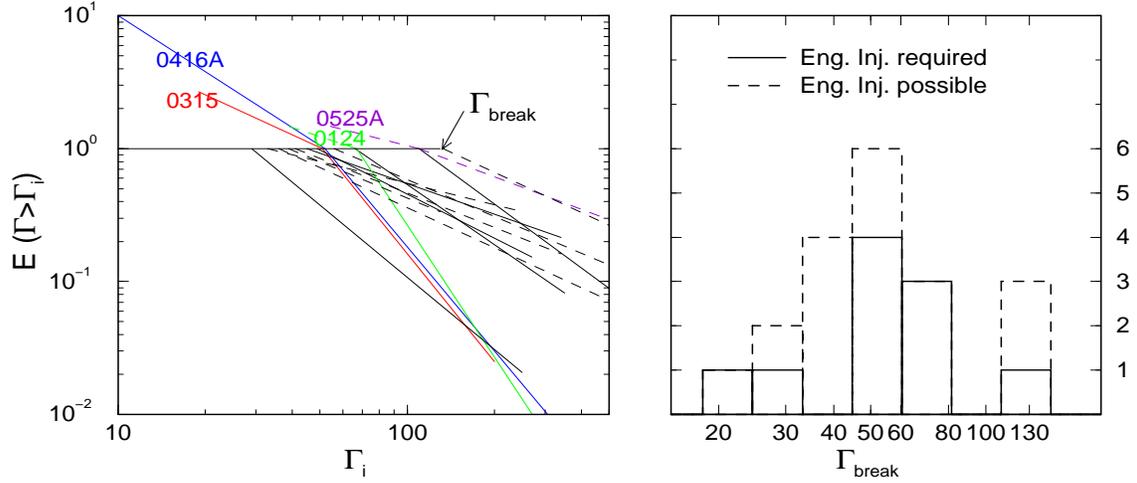}
  \caption{ Left panel: distribution of energy with ejecta Lorentz factor for afterglows with known 
      redshift. During the slow-decay phase, $\Gamma_i > \Gamma_{break}$, and the forward-shock energy 
      increases as more ejecta arrive at it. 
      Right panel: distribution of the cut-off Lorentz factor below which the incoming ejecta do not 
      carry significant energy (with the exception of the afterglows indicated in the left panel by
      their last four digits).
      Both panels: solid lines indicate afterglows for which energy injection is required by the 
      slow-decay phase; for dashed lines, energy injection is required if X-ray is below cooling,
      but not required if X-ray is above cooling. }
\label{ei}
\end{figure}

\subsection{Structured outflow}

 An analytical derivation of the decay index $\alpha$ of the emission from a structured outflow is
limited to an axially symmetric outflow and an observer located on the outflow symmetry axis.
Considering a power-law angular distribution of ejecta kinetic energy per solid angle, $\E \propto \theta^q$, 
with angle $\theta$ measured from the symmetry axis, and solving for the parameter $q$ which reconciles
the observables $\alpha_{x2}$ and $\beta_{x2}$, one reaches the same expressions as those given in 
equations (\ref{es0}) and (\ref{es2}) for the energy injection parameter $e$ and the same average values 
for the afterglows of figure \ref{x2} as those given in \S\ref{enginj} for $e$.

 In the more general case of an off-axis observer location, numerical calculations are needed to constrain 
the outflow parameters that yield a slow decay when the region carrying most of the outflow energy becomes 
visible. Eichler \& Granot (2006) have shown that the slow-decay phase of the afterglow 050315 can be
explained in this way, with a Gaussian jet seen off-axis. The light-curves obtained for such an outflow
structure and for various observer offsets $\theta_{offset}$ relative to the jet's axis are illustrated 
in figure \ref{so} (left panel). As shown there, for larger $\theta_{offsets}$, the emergence of the jet 
emission from the GRB tail (the fast-decay phase) occurs at a later time $t_{x2}$, a lower luminosity 
$L_{x2}$, and exhibits a smaller decay index $\alpha_{x2}$ during the slow-decay phase decreases. Therefore, 
this type of structured outflow model can be tested by searching for a $L_{x2}-\alpha_{x2}$ correlation 
and a $t_{x2}-\alpha_{x2}$ anticorrelation. A $L_{x2}-t_{x2}$ anticorrelation is not a relevant test because 
any model yielding decaying light-curves should produce it.

\begin{figure}
   \hspace*{-1cm} \parbox[h]{10cm}{ \includegraphics[width=8cm,height=8cm]{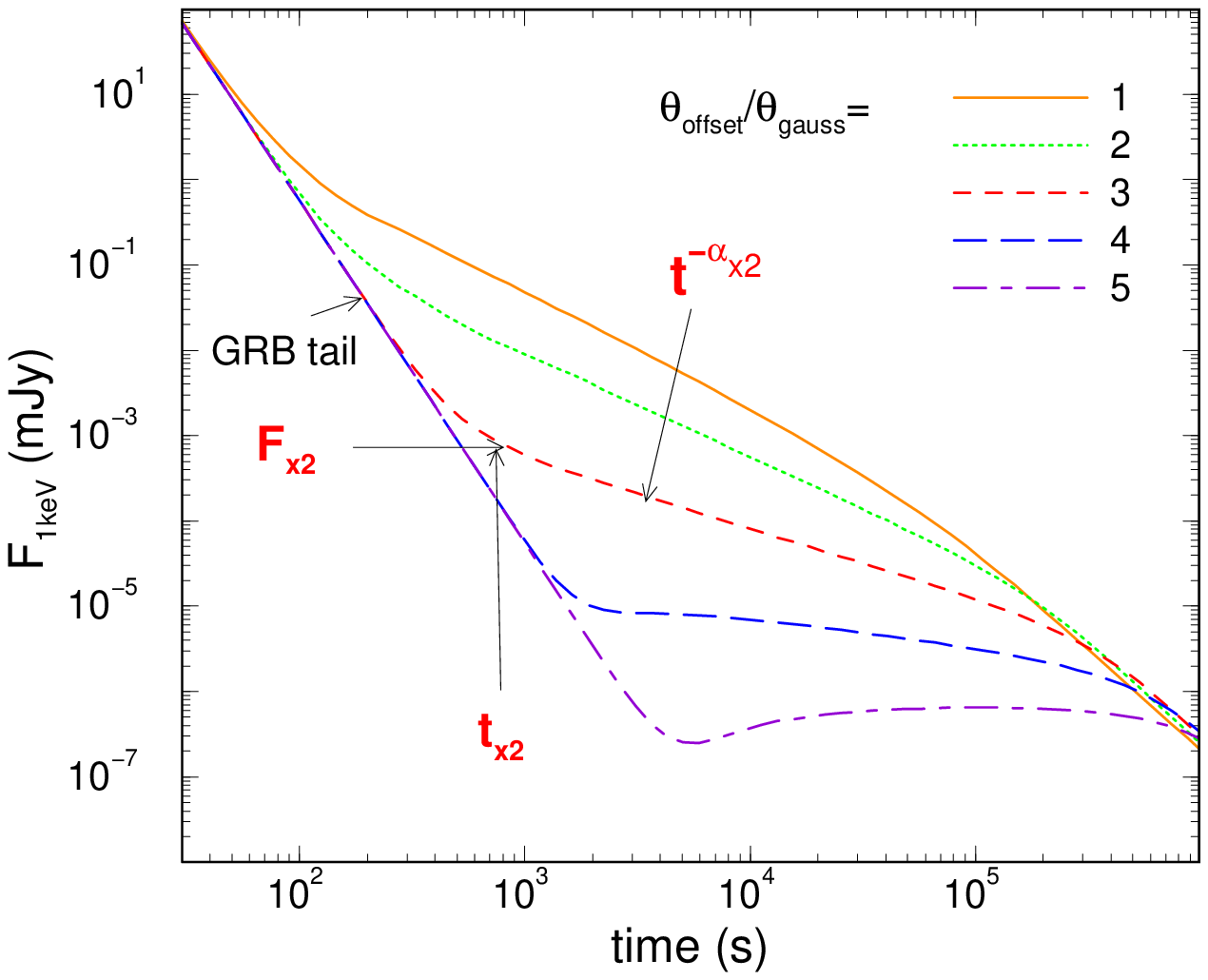} }
   \hspace*{-2cm} \parbox[h]{7cm}{ \includegraphics[width=7cm,height=8cm]{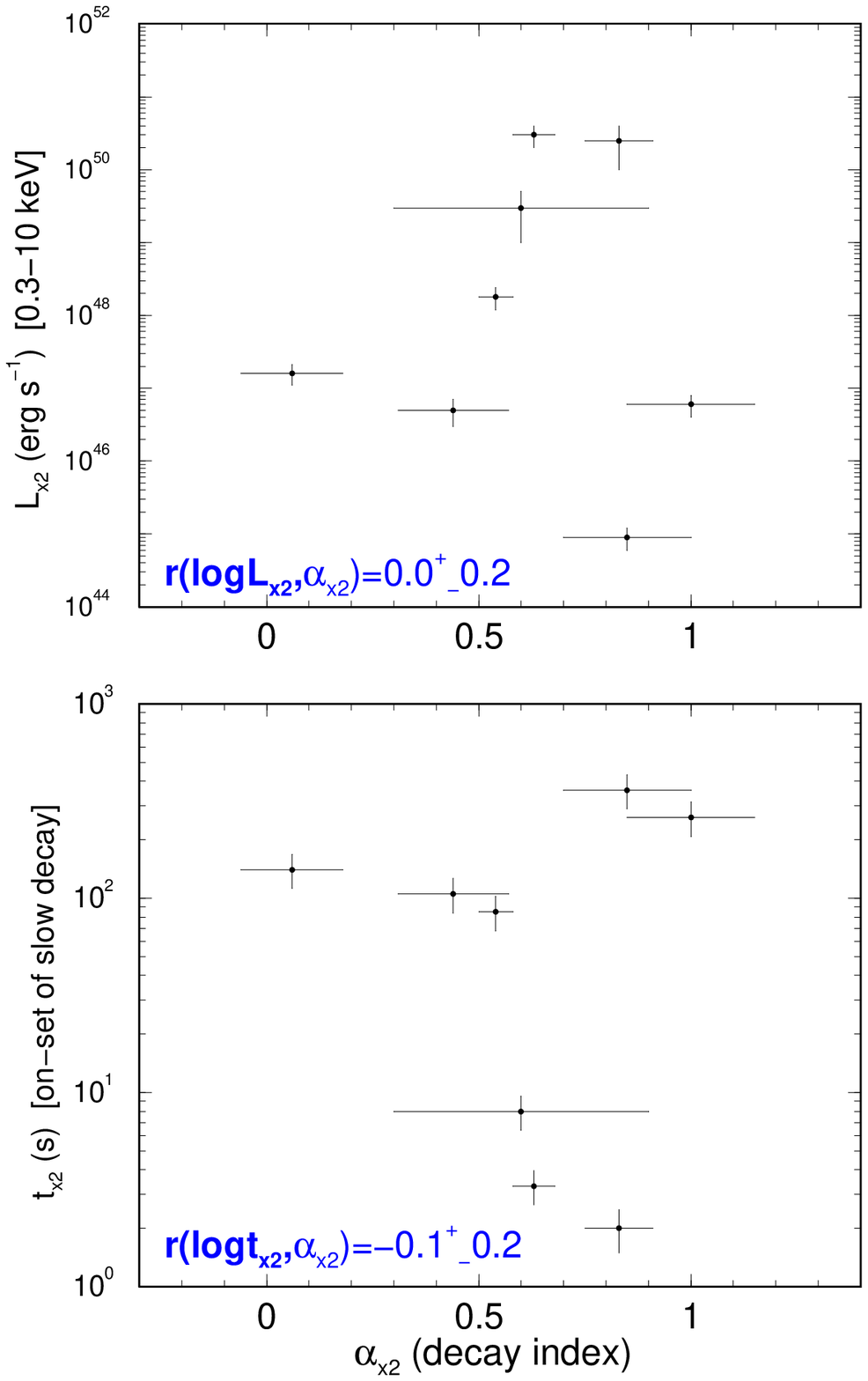} }
\caption{ Left panel: X-ray light-curves from a Gaussian jet seen off-axis emerge at $t_{x2}$ from under 
     the steeply falling-off GRB tail emission and can explain the slow-decay afterglow phase,
     $F_{\rm 1\;keV} \propto t^{-\alpha_{x2}}$. Legend gives the observer's angular offset, 
     $\theta_{offset}$, relative to the jet symmetry axis in units of the jet's Gaussian angular 
     scale, $\theta_{gauss}$. $L_{x2}$ is the k-corrected 0.3--10 keV luminosity.
   Right panels: test of the $L_{x2}-\alpha_{x2}$ correlation and $t_{x2}-\alpha_{x2}$ anticorrelation
     expected if the slow-decay phase arises from a structured outflow seen off-axis.  }
\label{so}
\end{figure}

 The right panels of figure \ref{so} show that the few afterglows with known redshifts (necessary for the
calculation of the afterglow intrinsic $L_{x2}$ and $t_{x2}$) and well-monitored transitions from the fast- 
to the to the slow-decay phases do not confirm the existence of the above correlations. Evidently, more 
afterglows are needed for a conclusive test of the structured, off-axis--seen outflow model for the slow-decay
phase.

\subsection{Evolving microphysical parameters}

 The afterglow decay also depends on the microphysical parameters $\epsi$ and $\epsB$ which quantify the 
fractional post-shock energy that is imparted to the relativistic electrons and magnetic field. The reason 
for this is that the afterglow X-ray flux depends on three characteristics of the forward-shock synchrotron 
spectrum (peak flux $F_p$, peak frequency $\nu_p$, and cooling frequency $\nu_c$ -- see figure \ref{spek}), 
whose evolutions depend on those of $\E$, $\epsi$, $\epsB$, on the stratification of the circumburst medium 
and on the geometry of the outflow. For the scalings of $F_p$, $\nu_p$ and $\nu_c$ with afterglow parameters 
given in figure \ref{spek} for a wind-like medium and an isotropic outflow, the resulting X-ray flux has the 
following scalings: 
$F_x \propto \E^{(p+1)/4} \epsi^{p-1} \epsB^{(p+1)/4}$ for X-ray below cooling and 
$F_x \propto \E^{(p+2)/4} \epsi^{p-1} \epsB^{(p-2)/4}$ for cooling below X-ray. 
Therefore, microphysical parameters increasing in time can lead to decays slower than expected in the 
standard forward-shock model. As discussed in \S\ref{evol}, optical light-curve decays in conjunction with
the X-ray observations of the slow-decay and pre jet-break phases require that microphysical parameters 
evolve in at least a few afterglows.

\begin{figure}
    \parbox[h]{9.5cm}{ \includegraphics[height=7cm,width=9.5cm]{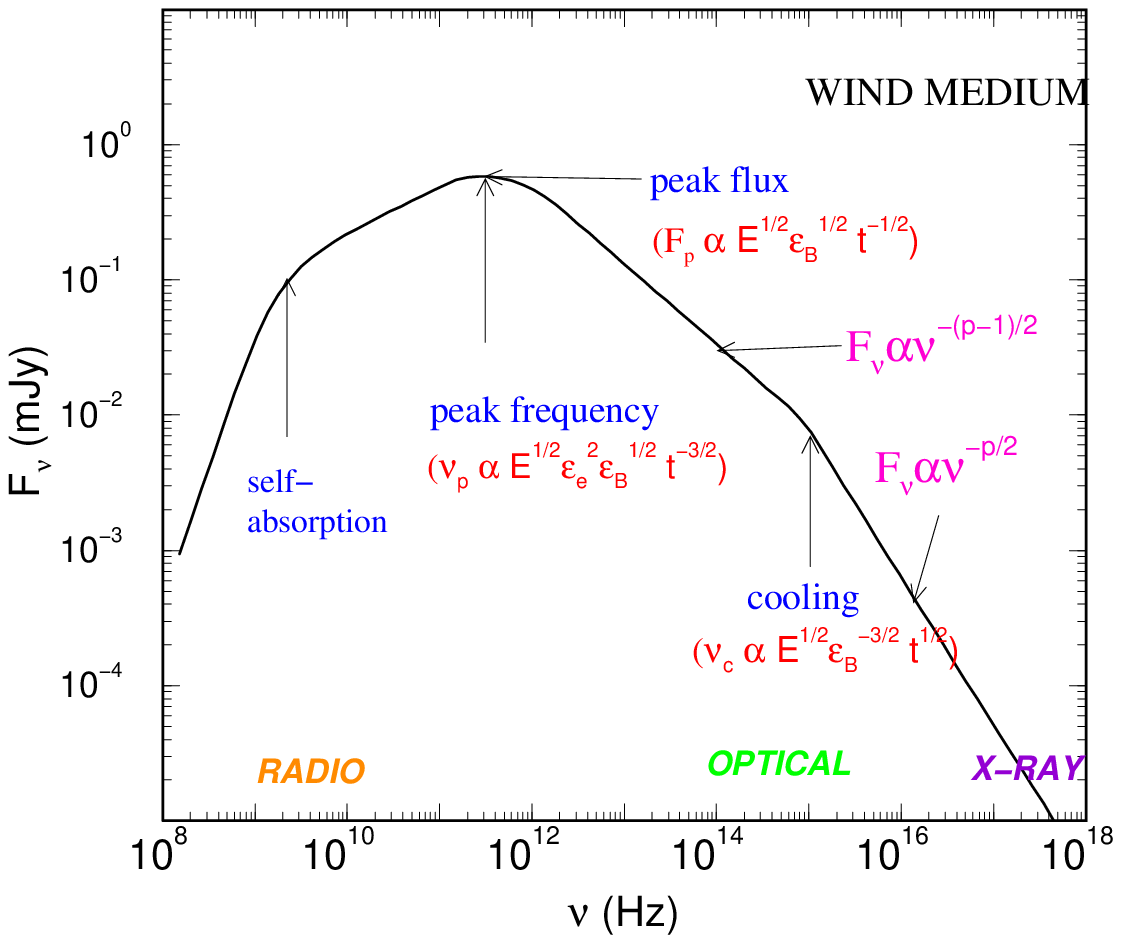} }
    \hspace*{2mm}  \parbox[h]{5cm}{ \footnotesize
      Spectrum of synchrotron emission from the forward shock, with its three break frequencies.
      The cooling frequency, $\nu_c$, is the synchrotron characteristic frequency for the electrons whose 
      radiative and adiabatic cooling timescales are equal. 
      Dependencies of the spectral characteristics on the afterglow model and their intrinsic temporal
      evolution are indicated for a wind-like medium ($n \propto r^{-2}$).}
\caption{}
\label{spek}
\end{figure}

\subsection{GRB efficiency}
\label{effi}

\begin{figure}
    \parbox[h]{9cm}{ \includegraphics[height=8cm,width=9cm]{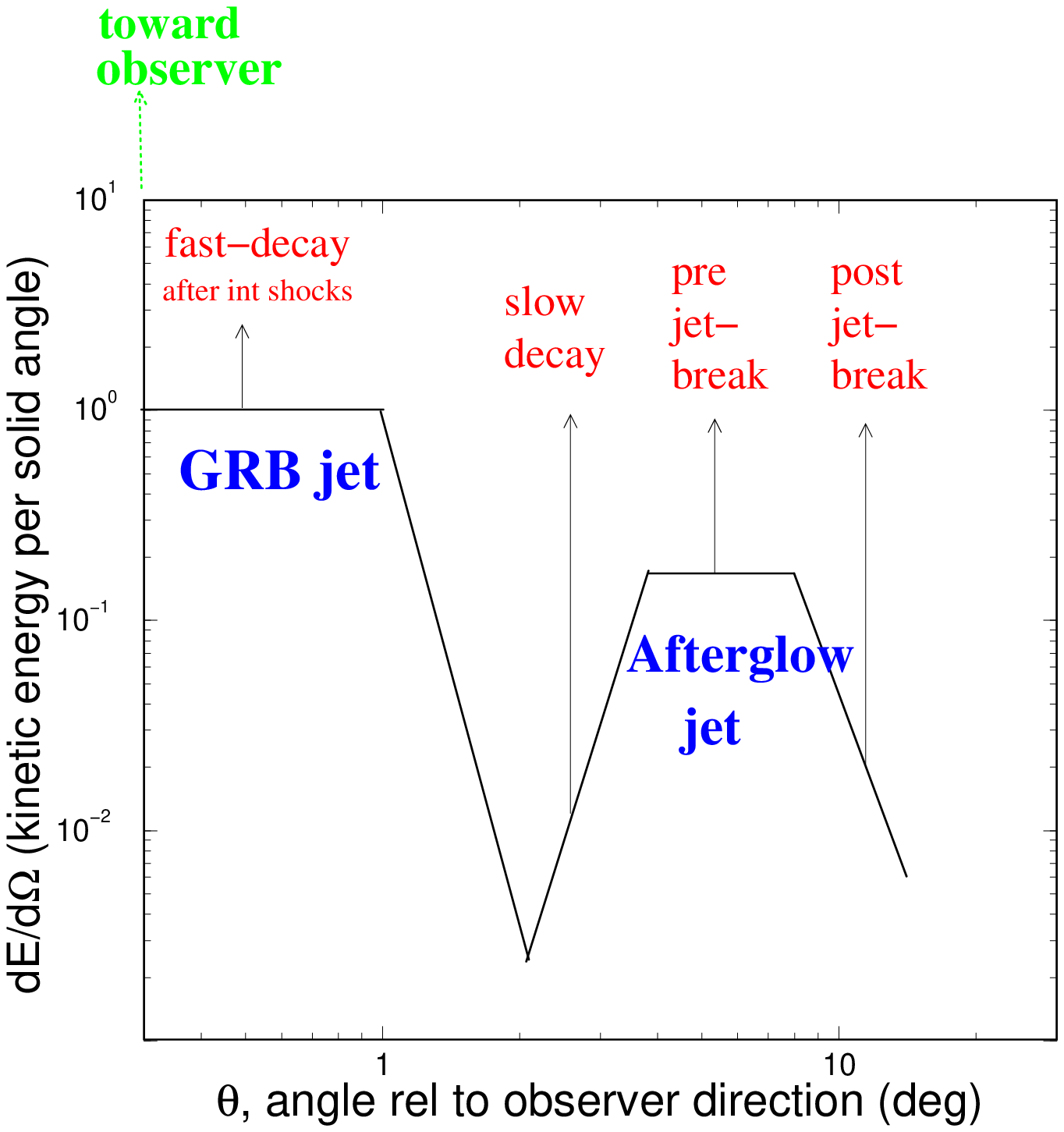} }
    \hspace*{2mm}  \parbox[h]{5cm}{ \footnotesize
        Energy distribution with angle in a longitudinal slice (defined by the afterglow-jet axis and
        the center-observer direction) through an outflow consisting of a narrow jet moving toward the
        observer (producing the GRB emission) and a wider jet seen off-axis (and whose radiation
        becomes dominant during the afterglow phase). X-ray afterglow phases are identified with the
        outflow region from where they arise.}
\caption{}
\label{dEdq}
\end{figure}
                                                                                                                   
 The ratio of the GRB output per solid angle, $\E_\gamma$, and the forward-shock kinetic energy per solid 
angle, $\E_{fs}$, determined either analytically (from the X-ray flux at $\sim 0.5$ d -- \eg Freedman \& Waxman 
2001, Lloyd-Ronning \& Zhang 2004, Fan \& Piran 2006, Granot, Konigl \& Piran 2006) or numerically (by 
modelling the broadband radio, optical, and X-ray 0.1--10 d afterglow emission -- Panaitescu \& Kumar 2002), 
is found to be 10\%--90\% for pre-Swift bursts and 1\%--30\% for Swift bursts. Such values pose a problem 
for the internal-shock model, for which an efficiency of $\sim 1$\% is expected (Kumar 1999, Spada, Panaitescu 
\& M\'esz\'aros 2000). However, the true 
efficiency of the burst mechanism is the ratio of $\E_\gamma$ to the ejecta kinetic energy per solid angle, 
$\E_{ejecta}$, which released the burst. Thus the true GRB efficiency can be smaller than that inferred 
using $\siml 1$d afterglow observations if $\E_{fs}$ underestimates $\E_{ejecta}$.

 The energy injection model for the slow-decay phase may lead to a higher true GRB efficiency if the ejecta 
which arrive at the forward shock after the burst did not participate in the production of $\gamma$-rays 
because, in this case, $\E_{ejecta}$ is less (by a factor 10--100) than $\E_{fs}$. This disfavours a 
long-lived engine and favours a short-lived one because, in the former model, most of the forward-shock 
energy at 1 d is from ejecta which were released after the burst. 

 The structured outflow model has the potential of lowering the true GRB efficiency if the $\E_{ejecta}$
of the outflow visible during the GRB phase (the $\theta \siml 1\deg$ GRB jet) is larger than the average 
$\E_{fs}$ of the region visible at $t \siml 1$ d (\ie within an opening $\theta < [\Gamma({\rm 1d)}]^{-1}$ 
of several degrees). A second condition to be satisfied is that the emission from the wider, afterglow outflow 
dominates that from the GRB jet during the afterglow phase, otherwise the afterglow and GRB fluxes would be 
both a measure of the kinetic energy of same ejecta (the GRB jet) and the true GRB efficiency would be the 
same as that inferred from afterglow observations. Figure \ref{dEdq} shows schematically such a dual outflow.

 If microphysical parameters decrease in time then $(i)$ a higher fraction of the shock energy is imparted 
to electrons during the burst and $(ii)$ the cooling frequency during the burst is lower than during the 
afterglow phase. Either factor can increase the burst radiative efficiency, hence the evolving microphysical 
parameters model may also explain in part the high apparent GRB efficiency.

\section{Pre jet-break phase}
\label{pre}

\hspace*{3mm} 
 The definition for this phase is that it is either preceded by a slower decay or followed by a steeper one 
or, if observations of these adjacent phases lacks, that it begins later than 1 hour after the trigger. 
The latter condition avoids confounding the pre jet-break phase of poorly monitored afterglows with the 
slow-decay phase. In some afterglows this phase ends at $\siml 1$ d, being followed by a steeper decay, 
while in others it lasts for more than 10 d.

 Figure \ref{x3} shows that, for 27 afterglows, the X-ray decay index $\alpha_{x3}$ and the spectral slope 
$\beta_{x3}$ during it are consistent with the predictions of the standard forward-shock model for a spherical 
outflow, \ie at times when $\Gamma > \theta_{jet}^{-1}$ ($\theta_{jet}$ being the half-angle of the outflow) 
and the jet boundary is not yet visible to the observer. This suggests that, during the pre jet-break phase, 
the basic assumptions (\S\ref{in}) of the standard forward-shock model (constant shock energy, uniform
angular distribution of the ejecta kinetic energy, constant microphysical parameters) may be valid. However, 
the $\alpha_{x3}-\beta_{x3}$ consistency does not show those assumptions to be clearly satisfied because the 
indices/slopes uncertainties may be sufficiently large to reach an accord with one of the three possible 
models shown in figure \ref{x3}.
 
\begin{figure}
    \parbox[h]{8.5cm}{ \includegraphics[height=7cm,width=8.5cm]{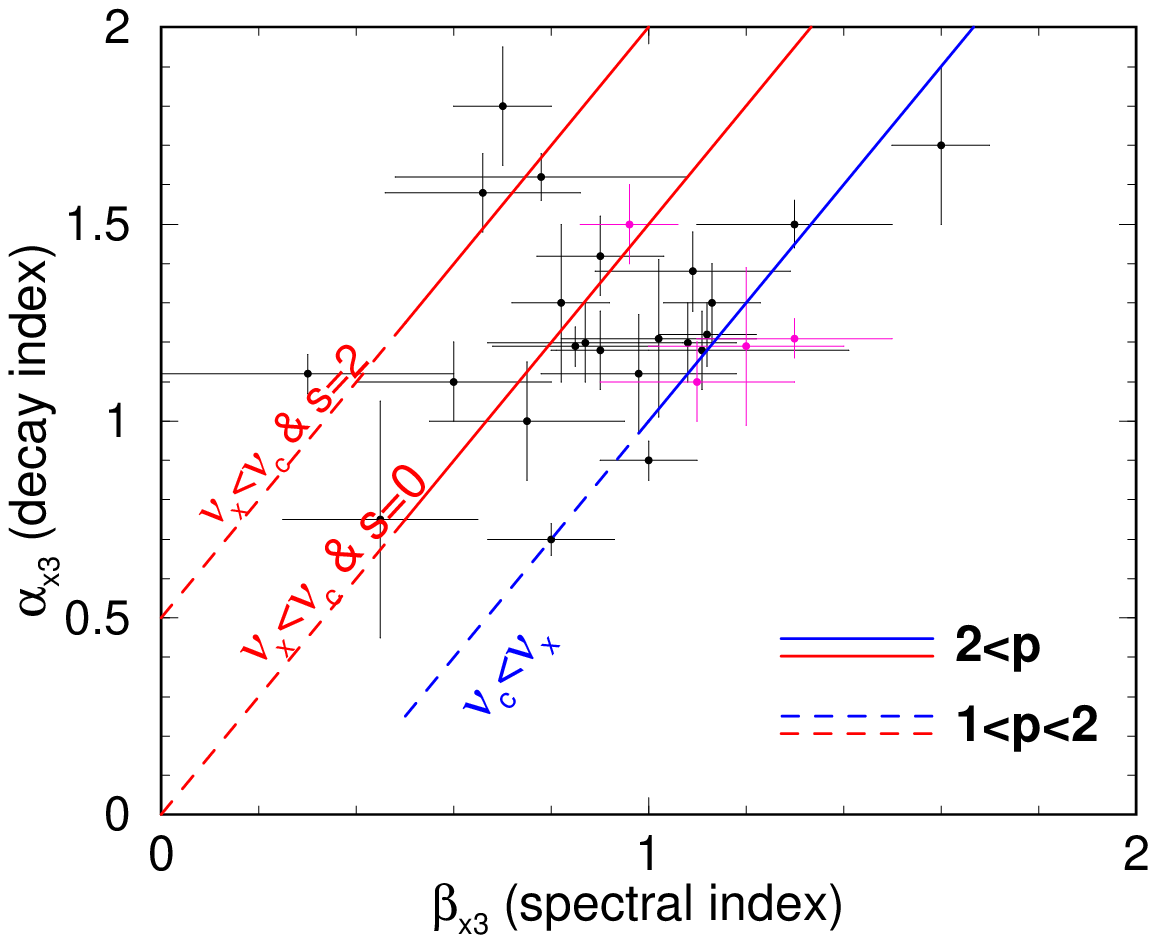} }
    \hspace*{2mm}  \parbox[h]{6cm}{ \footnotesize
        Comparison between decay index \& spectral slope during the pre jet-break phase and the
        expectations of the standard forward-shock model. }
\caption{}
\label{x3}
\end{figure}

 That no significant change in the X-ray spectral slope is measured at $t_b \sim 1$ h (Nousek \etal 2006),
when the slow-decay phase ends and the pre jet-break phase begins, shows that the X-ray light-curve steepening 
at $t_b$ is not caused by the passage of a spectral break through the X-rays. Taking into account the 
three mechanisms discussed in \S\ref{slow} for the slow-decay phase, the consistency of $\alpha_{x3}$ and
$\beta_{x3}$ of the pre jet-break phase with the standard forward-shock model suggests that the X-ray
break at $t_b$ should be identified with $(i)$ cessation of energy injection at $t_b$, or $(ii)$ a structured 
outflow whose region of maximal $\E$ becomes visible at $t_b$, or $(iii)$ $\epsi$ and $\epsB$ becoming constant 
after $t_b$.

\subsection{Chromatic X-ray breaks and evolving microphysical parameters}
\label{evol}

 All three mechanisms above for the X-ray breaks seen at $t_b \sim 1$ h entail a change in the evolution of 
the flux and characteristic frequencies of the forward-shock synchrotron spectrum (figure \ref{spek}),
hence they all should lead to an achromatic light-curve break, appearing at $t_b$ in the afterglow light-curve
at any frequency. However, as illustrated in figure \ref{xbrk}, an optical light-curve break is quite often 
not seen at the end of the X-ray slow-decay phase. 
  
\begin{figure}
  \includegraphics[width=15cm,height=14cm]{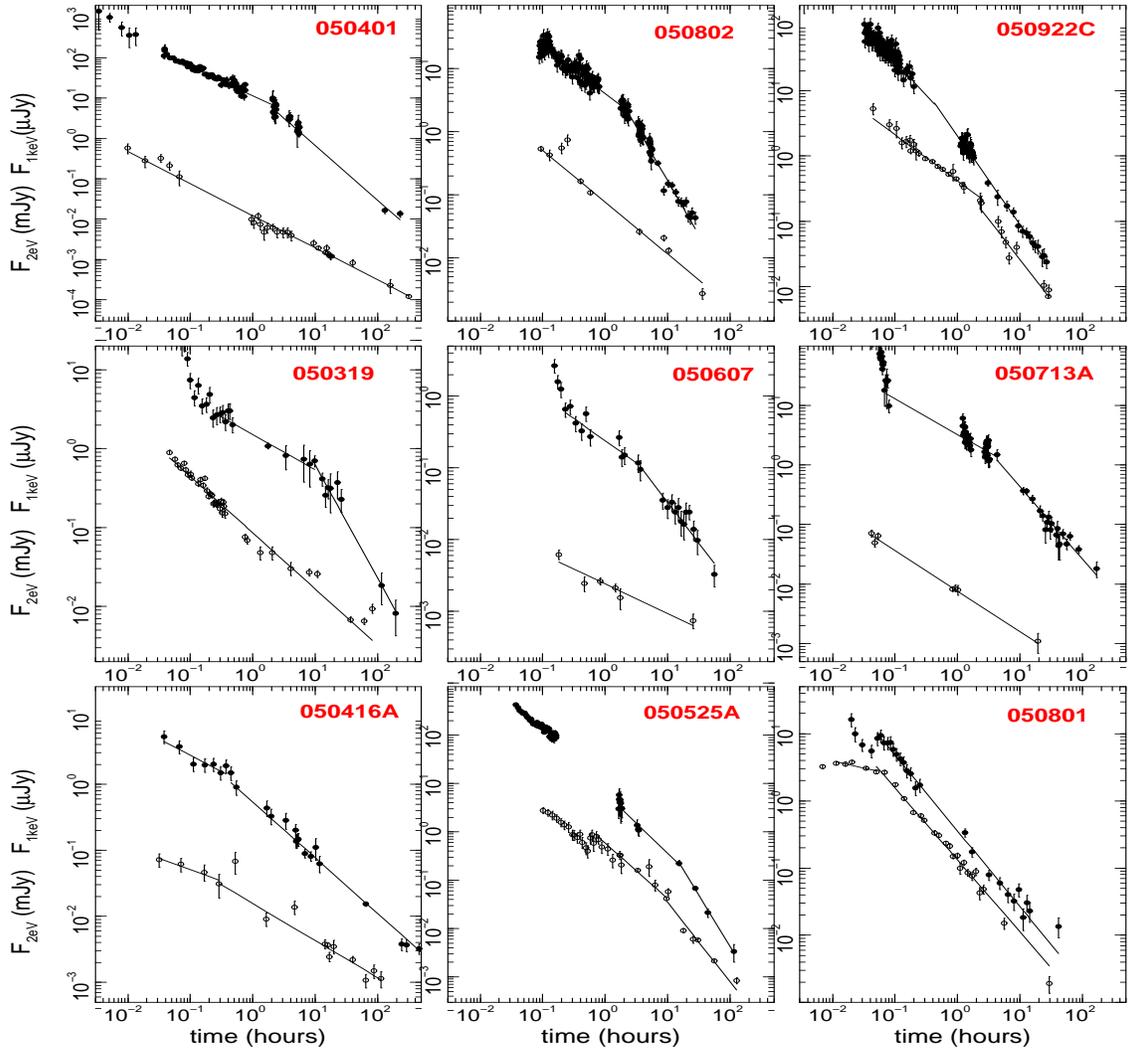}
  \caption{ Compilation of Swift afterglows with good coverage in both X-ray (filled symbols, upper light-curve) 
    and optical (open symbols, lower light-curve). The afterglows in top panels exhibit a chromatic X-ray break 
    at the end of the slow-decay phase (\ie the break is not seen in the optical as well). A chromatic X-ray 
    break is possible for the afterglows in the middle panel, but optical monitoring after the X-ray break is 
    insufficient to prove it. The afterglows in the bottom panels exhibit an achromatic break (\ie seen 
    simultaneously in the optical and X-ray). Given the steep post-break decay, the break of 050525A is 
    more likely a jet-break. For most afterglows, the pre-break optical and X-ray decay indices are comparable.}
\label{xbrk}
\end{figure}

 The decoupling of the optical and X-ray decays during the pre jet-break phase implies the existence of a spectral 
break between the optical and X-ray, whose evolution leads to a different decay index of the emission at these 
two frequencies. That spectral break could arise from the shock-acceleration of electrons but then it would be 
surprising that, over 2--4 decades in observer time, that break does not cross the optical domain. More likely, 
the break frequency in question is the cooling frequency, $\nu_c$. Given that the afterglow spectrum steepens
by $\delta \beta = 1/2$ at $\nu_c$, the observed optical and X-ray and decay indices require that the evolution
of $\nu_c$ before and after the X-ray break (at $t_b$) is $d\ln\nu_c/d\ln t=2(\alpha_{o2/o3}-\alpha_{x2/x3})$.
  
 For the six afterglows in the top and middle panels of figure \ref{xbrk}, which display a chromatic X-ray 
break, the required evolution of $\nu_c$ at $t < t_b$ ranges from $t^{-1/2}$ to $t^{1/2}$, being consistent with 
no evolution for four of them. The required evolution of $\nu_c$ at $t > t_b$ ranges from $t^{-1}$ to $t^{-7/4}$.
Such a fast evolution of $\nu_c$ is marginally problematic only for GRB 050401 for which the unbroken 
optical power-law decay continues for 2 decades in time after $t_b$. During this time, $\nu_c$ would cross the 
optical domain at the epoch of the last optical measurements (10 d), if it were just under the X-ray domain at 
$t_b$. 
 
 It can be shown that the expected cooling frequency evolution is $\nu_c \propto \E^{-1/2} \epsB^{-3/2} t^{-1/2}$ 
for a homogeneous medium (Sari, Piran \& Narayan 1998, Panaitescu \& Kumar 2000) and $\nu_c \propto \E^{1/2} 
\epsB^{-3/2} t^{1/2}$ for a wind (Chevalier \& Li 1999, Panaitescu \& Kumar 2000).  Energy injection can
increase $\E$ while radiative losses decrease it as $t^{-3/7}$ for a homogeneous medium and as $t^{-1/3}$ for
a wind (Panaitescu \& Kumar 2004). Then, the required evolutions of $\nu_c$ imply that $\epsB$ must evolve at 
$t < t_b$, for either type of medium, and that it evolves also at $t > t_b$ for a wind-like medium.
As a side remark, this shows that the above-mentioned consistency of the decay index and spectral slope 
during the pre jet-break phase with the standard forward-shock model expectations is, sometimes, fortuitous. 

  It is tempting to attribute the X-ray break to a change in the evolution of $\epsB$. However, as discussed 
above, no afterglow parameter acting alone can produce a chromatic light-curve break. The natural next step 
is to assume that $\epsi$ also evolves and to require that its evolution changes at $t_b$ such that it hides 
in the optical the break that the evolution of $\epsB$ would produce if it acted alone. This scenario is twice 
contrived. One odd feature is the fine-tuning of the changes in the evolutions of $\epsi$ and $\epsB$ required
to "iron-out" the optical break. Second is that, as long as the dynamics of the outflow is unaltered, a change 
in the evolution of microphysical parameters lacks a reason.
 
 A less contrived explanation for the chromatic X-ray breaks should be based on a change in the outflow dynamics
at the time $t_b$ of the break. Such a change would naturally occur from the cessation of energy injection
at $t_b$: $d\E/d\Gamma (t<t_b) \propto \Gamma^{-(e+1)}$ and $d\E/d\Gamma (t>t_b) = 0$. 
This model would be less contrived if microphysical parameters were dependent on the shock's Lorentz factor 
$\Gamma$ as, in this case, the changing evolution of $\epsB$ at $t_b$ would be tied to the change in the outflow 
dynamics at the end of energy injection. Furthermore, if $\epsi$ also evolved, then it would be possible to explain
the chromatic X-ray breaks with a model where the evolution of microphysical parameters with $\Gamma$ is steady 
across $t_b$, \ie their dependence on $\Gamma$ before and after $t_b$ is the same and only the evolution
of $\Gamma$ changes at $t_b$. Below, we investigate this model.

 That optical and X-ray light-curves exhibit power-law decays implies that the evolving $\epsi$ and $\epsB$ are 
power-laws of $\Gamma$: $\epsi \propto \Gamma^{-i}$, $\epsB \propto \Gamma^{-b}$. The exponents $e$, $i$, and $b$ 
can then be determined from the observed optical and X-ray decay indices. Because these indices provide four 
constraints, we shall also allow a free stratification of the ambient medium density: $n(r) \propto r^{-s}$.
As argued above, to decouple the optical and X-ray decays requires that cooling is between optical and X-ray, 
therefore the index of the electron distribution with energy is $p = 2\beta_{x2}$.

 In this modified forward-shock model, the resulting decay indices for the slow-decay phase are (Panaitescu \etal 
2006) :
\begin{equation}
 \alpha_{o2} = \frac{s}{8-2s} + \frac{3}{4} (2\beta_{x2}-1) - \frac{3-s}{e+8-2s}  \left[
        \left(\frac{2\beta_{x2}+3}{4}-\frac{s}{8-2s}\right) e + (2\beta_{x2}-1)i + \frac{2\beta_{x2}+1}{4}b \right]
\end{equation}
\begin{equation}
 \alpha_{x2}  =  \frac{3\beta_{x2} - 1}{2} - \frac{3-s}{e+8-2s} \left[ 
       \frac{1}{2}(\beta_{x2}+1) e + (2\beta_{x2}-1)i + \frac{1}{2}(\beta_{x2}-1)b \right] \,.
\end{equation}
The decay indices for the pre jet-break phase are obtained by setting $e=0$ in the above equations.

 After some calculations, it can be shown that the lack of an optical break ($\alpha_{o2} = \alpha_{o3}$) 
requires that  
\begin{equation}
  (2\beta_{x2}-1)i + \frac{1}{4}(2\beta_{x2}+1)b = 4\beta_{x2} - 2\alpha_{o2}  \quad {\rm and} \quad
   s = 4 \frac{\alpha_{o2}+3}{2\beta_{x2}+5} \;.
\label{ib}
\end{equation}
The first equation above contains all that is contrived about this model.
For the six afterglows with chromatic X-ray breaks shown in figure \ref{xbrk}, $\overline{\beta_{x2}} = 
1.00 \pm 0.15$ and $\overline{\alpha_{o2}} = 0.69\pm0.15$, thus the condition for lack of an optical break 
at the end of energy injection (equation \ref{ib}) can be generalized as $i/3+b/4 \simeq 0.9$. 

 The second equation (\ref{ib}) leads to $s$ ranging from 1.9 to 2.3, being consistent with $s=2$ for five 
afterglows, \ie the circumburst medium has a wind-like stratification. The magnitude of the X-ray break, 
$\alpha_{x3} - \alpha_{x2}$, determines the energy injection index $e$, which we find to be between 2 and 4. 
The $i$ and $b$ indices for the evolution of $\epsi$ and $\epsB$ with $\Gamma$ are in the (-4,0) and (4,7) 
intervals, respectively. During the slow-decay phase, energy injection mitigates the outflow deceleration 
from the $\Gamma^{-1/2}$ expected for the no-injection case to $\Gamma^{-0.12\pm0.04}$.
 
 The evolution of microphysical parameters with observer time is $\epsB \propto t^{0.6\pm0.2}$ and 
$\epsi \propto t^{-0.2\pm0.2}$ during the slow-decay phase, $\epsB \propto t^{1.3\pm0.2}$ and
$\epsi \propto t^{-0.4\pm0.4}$ during the pre jet-break phase. If these evolutions start at the end of 
the burst then, for the six afterglows with chromatic X-ray breaks, $\epsi$ decreases by a factor 2--6 
while $\epsB$ increases by a factor 10--50 from the burst epoch to the end of the slow-decay phase. 
The decreasing fraction of shock energy imparted to electrons helps understanding the apparent, high GRB 
efficiency. The increasing fraction of shock energy in magnetic field leads to a cooling frequency during
the burst phase which is $\sim 100$ times larger than for a constant $\epsB$, which may not affect
the GRB efficiency because, during the burst, electrons are likely to cool on a timescale much shorter 
than the hydrodynamic timescale.

\section{Post jet-break phase}

\begin{figure}
    \parbox[h]{8cm}{ \includegraphics[height=7cm,width=8cm]{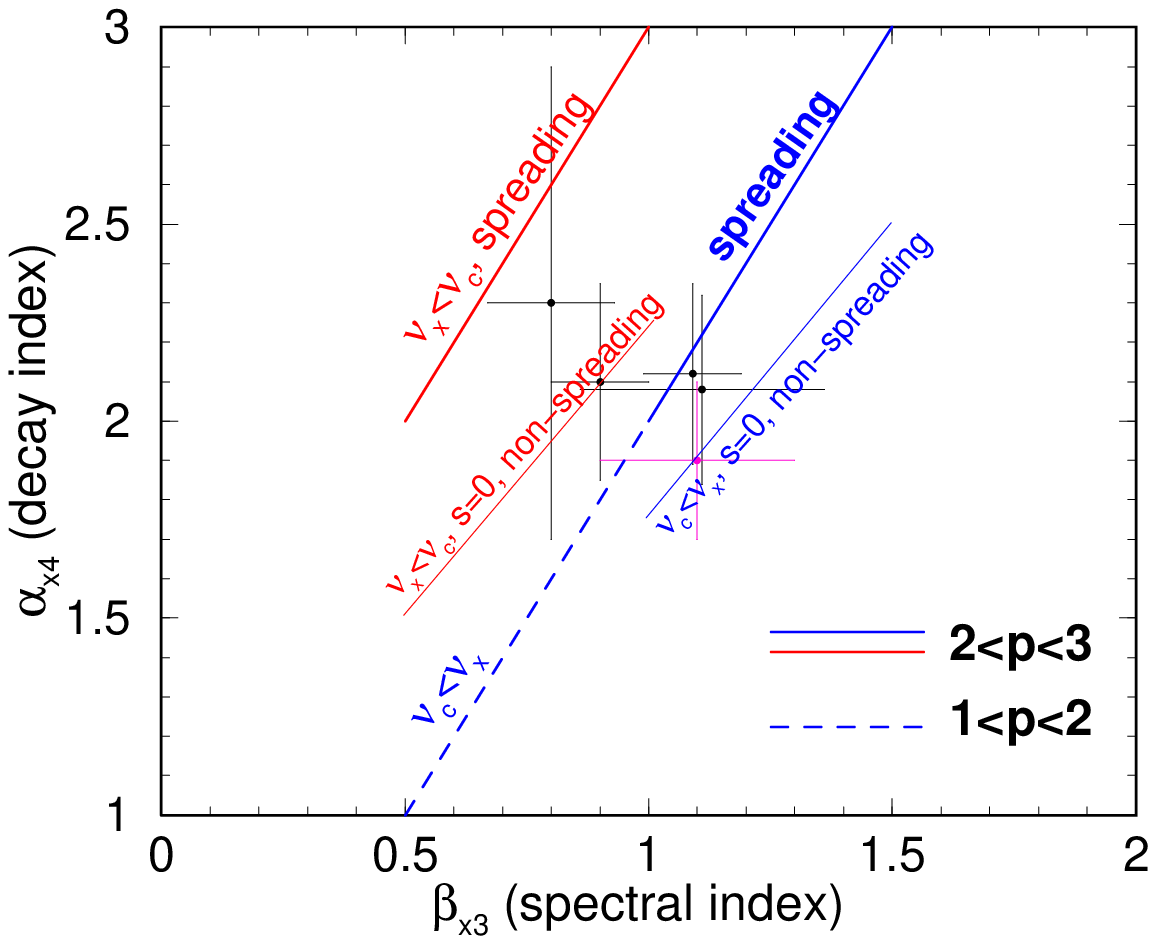} }
    \hspace*{2mm}  \parbox[h]{6cm}{ \footnotesize
       Comparison between decay index \& spectral slope during the post jet-break phase and the
       expectations of the standard forward-shock model. Thick lines are for a laterally spreading jet 
       (and any stratification of the ambient medium), thin lines are for a non-spreading jet and a
       homogeneous medium ( wind-like medium provides a poorer match and is not shown).  }
\caption{}
\label{x4}
\end{figure}

\hspace*{3mm} 
 For three afterglows, Swift has measured within the first day after trigger a steepening of the X-ray 
light-curve to $\alpha_{x4} \simg 2.0$, while two other afterglows exhibit a similar steep decay after 
about 3 d. A few other Swift afterglows do not display such a steepening until days or weeks after 
the burst, most notable being GRB 050416A, whose pre jet-break phase extends up to at least 40 d.

 As shown in figure \ref{x4}, for the few X-ray afterglows with a $\sim 1$ d break, the decay index and
spectral slope during this phase are consistent with the standard forward-shock emission from a jet whose 
boundary has become visible (\ie $\Gamma < \theta_{jet}^{-1}$). If the jet is spreading sideways at the 
sound speed then all cases are consistent with cooling being below X-ray. A model where the jet lateral 
expansion is impeded (conical jet) can also accommodate the observations, provided that the ambient medium 
is homogeneous. Consistency of the decay indices and spectral slopes with the standard jet model expectations 
suggests that its assumptions are correct. However, given their large uncertainties and the possible 
variants of the jet model, that is not a strong proof that the observed breaks are indeed due to the 
outflow collimation. A more stringent test of the standard jet model is consistency between the optical 
and X-ray decay indices before and after the break.  Further, the jet origin of the light-curve breaks 
could be tested through the expected closure relationships between the decay indices and spectral slopes, 
if optical and X-ray spectral slopes can be determined sufficiently well.

\begin{figure}
  \includegraphics[width=15.5cm,height=7.5cm]{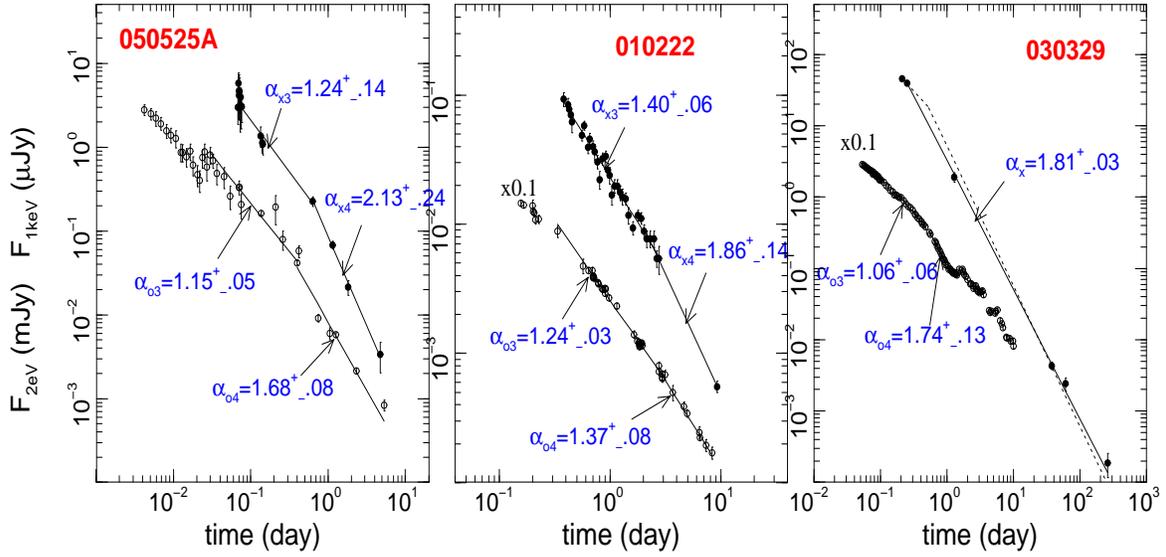}
  \caption{ Optical and X-ray light-curves for the afterglows with the best monitoring before and after
      the jet break. Filled symbols for X-ray, open symbols for optical. Decay indices of power-law
      fits to the pre and post jet-break decays are indicated. Simultaneous optical and X-ray breaks
      can be identified for all three afterglows. For 050525A (Swift burst) the pre jet-break decay indices
      are consistent but the post jet-break optical decay is too slow. For 010222, the X-ray break appears
      sharper than in the optical. For 030329, the pre jet-break X-ray decay is not well constrained,
      but the optical and X-ray decays after the jet-break at $t_b=0.5$ d are consistent with the jet model.
      The optical emission of 030229 after 1 day exhibits fluctuations and was not included in the fit.
      Dotted lines indicate an X-ray break consistent with that seen in the optical at 0.5 d. }
\label{jet}
\end{figure}

 For constant microphysical parameters, the pre jet-break optical and decay indices ($\alpha_{o3}$ and
$\alpha_{x3}$) should be equal (in this case, cooling frequency is not between optical and X-ray and 
$\beta_o = \beta_x$) or differ by 1/4 (for cooling frequency between optical and X-ray, implying 
$\beta_x - \beta_o =1/2$). The post jet-break indices ($\alpha_{o4}$ and $\alpha_{x4}$) should be equal 
if the jet is spreading laterally or differ by at most 1/4 if the jet is not spreading. 
In contrast, for light-curve breaks arising from a sudden change in the evolution of 
microphysical parameters, the index difference can be larger (provided that the cooling frequency is 
between optical and X-rays). Simultaneity of the optical and X-ray breaks is, evidently, a pre-requisite 
for a jet-break, however is not a smoking gun of it, as other mechanisms are also likely to yield an 
achromatic break (achromaticity rules out only light-curve breaks arising from the passage of a spectral 
break).

 The optical breaks of pre-Swift afterglows have been well monitored, but a sufficiently good temporal coverage 
before and after the jet-break was not achieved also in the X-rays. Even for the afterglows with the widest 
temporal coverage, GRB 010222 and GRB 030329 (figure \ref{jet}), the existence of an X-ray break simultaneous 
with the optical is not well-proven: that of 010222 is based on a single, 10 d Chandra measurement, while the 
sparsely monitored X-ray emission of 030329, extending over 3 decades in time, is consistent with a single 
power-law decay. Furthermore, the post-break X-ray decay of 010222 is too fast compared to that seen in the 
optical and is inconsistent with the standard jet model and constant microphysical parameters. On the positive
side, as shown by the dotted lines in figure \ref{jet}, the X-ray measurements of 030329 are compatible with
a jet-break occurring simultaneously with that in the optical, and may have the same pre and post jet-break 
decay indices as the optical light-curve, but they do not offer conclusive evidence for a jet-break.

 The available observations of Swift afterglows allow the jet-break test to be done for only three cases. 

 For GRB 050525A (figure \ref{jet}), both the optical and X-ray exhibit a break at $\sim 10$ h.
The difference between the post jet-break X-ray and optical decay indices, $\alpha_{x4} - \alpha_{o4} = 
0.49 \pm 0.25$, is at $2\sigma$ from the spreading-jet model expectation and at least $1\sigma$ above that 
expected for a non-spreading jet. Hence, the problem with the afterglow 050525A is that its post jet-break 
optical decay is too slow. We note that there is no evidence of a contribution from the host galaxy that 
could account for the too slow optical and that the associated supernova dominates the afterglow flux only 
after 3 d (Della Valle \etal 2006). That $\alpha_{x4} - \alpha_{o4} > 1/4$ requires evolving microphysical 
parameters and cooling frequency to be between optical and X-ray. The latter requirement is marginally 
consistent with the pre-break index difference, $\alpha_{x3} - \alpha_{o3} = 0.09 \pm 0.15$ (which is at 
$1\sigma$ below the expected value) but the former requirement is at somewhat at odds with the constancy
of microphysical parameters before the jet-break indicated by the near equality of $\alpha_{x3}$ 
and $\alpha_{o3}$. Thus the slow post-break optical decay of GRB afterglow 050525A can be accommodated by
a jet only if microphysical parameters start evolving at about the jet-break epoch.

 The optical and X-ray light-curves of GRB afterglow 060124 (Curran \etal 2006) display a break 
at $\sim 18$ h, with $\alpha_{x3} - \alpha_{o3} \simeq 0.27$ and $\alpha_{x4} - \alpha_{o4} \simeq 0.33$. 
The former index difference is consistent with the standard forward-shock model expectations for a homogeneous
medium and cooling frequency between optical and X-ray. The latter index difference implies that the jet does 
not expand laterally, which also explains the rather slow post-break decay ($\alpha_{x4} \sim 1.7$).
Furthermore, a homogeneous medium and $\nu_o < \nu_c < \nu_x$ are also required for the optical pre-break decay 
index to be consistent with the optical spectral slope. Lastly, $\nu_o < \nu_c < \nu_x$ is in accord with the 
difference between the optical and X-ray spectral slopes: $\beta_{x3} - \beta_{o3} \simeq 0.6$. Thus, the break 
at 0.75 d of the GRB afterglow 060124 is fully consistent with a jet origin.

 Simultaneous optical and X-ray breaks are observed for GRB afterglow 060526 at $\sim 35$ h. 
The optical and X-ray decay indices are equal both before and after the break, indicating that cooling 
frequency is above X-ray and that the jet expands laterally. The latter is consistent with the steepness 
of the post-break decay: $\alpha_{x4} \simeq 2.8$.

\section{Summary}

\hspace*{3mm} 
{\bf Fast-decay phase}.
 The temporal and spectral properties of the fast-decaying emission are consistent with it being the 
large-angle GRB emission. For this model, decays faster than expected result when the origin of time 
for the last GRB pulse (which dominates the tail emission over a two-fold time increase) is well after
trigger. Slower-than-expected decays, as seen in two short GRBs, could be due to that the fast-decay 
phase is dominated by the forward shock. Structure of the ejecta emission on an angular scale of 
$\Gamma^{-1}$ could also explain both departures from expectations and may also account for the 
fast-decay emission being softer than that of the burst. 

{\bf Slow-decay phase}.
 This and the following phases can be attributed to the forward shock which energizes the circumburst 
medium. Various mechanisms can account for the slow decay: energy injection in the shock, a structured 
outflow where ejecta kinetic energy per solid angle increases with angle measured from the center--observer 
axis, or evolving microphysical parameters.

 In the energy injection model, a cumulative energy of the incoming ejecta increasing as $\Gamma^{-1}$ to 
$ \Gamma^{-3}$, down to $\Gamma_{break} = 50 \pm 25$ is required, though for a few afterglows there could 
be a dynamically-significant energy in slower ejecta as well.
A long-lived engine, releasing ejecta over a lab-frame duration comparable to the observer-frame time 
when the slow-decay is observed, exacerbates the high GRB efficiency (10\%-90\%) issue, as in this model 
most of the relativistic ejecta were released after the burst and did not contribute to the GRB emission. 
Hence, the high efficiency at which the ejecta kinetic energy is converted into $\gamma$-rays favours a 
short-lived engine, where all the ejecta are released simultaneously, contribute to the GRB emission, and 
undergo a delayed interaction with the forward shock due to the spread (by a factor of several) in their 
Lorentz factor.

 A structured outflow whose region of maximum energy per solid angle becomes gradually visible to the
observer can also account for the slow-decay phase. However, for the limited sample of afterglows with 
redshifts and well-observed transitions to the slow-decay phase, the correlations expected in this model
among the X-ray luminosity, decay index, and epoch of the slow-decay are not confirmed. This model may 
explain the apparent, high GRB efficiency, provided that the core of the outflow which releases the burst 
emission has a higher kinetic energy per solid angle than the average for the outflow visible during 
the afterglow phase, but does not overshine it during the afterglow phase.

{\bf Pre jet-break phase}.
 The temporal and spectral properties of the X-ray emission during this phase phase are consistent 
with the standard forward-shock model expectations at times when the shock is sufficiently relativistic 
that the region visible to the observer is smaller than the opening of the outflow. This would suggest 
that the shock energy and microphysical parameters are constant. However, the chromatic X-ray light-curve 
breaks at the end of the slow-decay phase (which are not accompanied by a simultaneous optical break) 
indicate that, if the optical and X-ray emissions arise from the same outflow, then microphysical 
parameters are not constant.

  A sudden change in the evolution of microphysical parameters could explain the chromatic X-ray breaks 
but that would require the pre- and post-break evolutions to be correlated so that they "iron out" the 
optical light-curve break. Furthermore, without a change in the forward shock dynamics, that sudden 
change in the evolution of microphysical parameters has no physical motivation. A less contrived 
model is that where the X-ray break is caused by a change in the outflow dynamics, due to cessation 
of energy injection in the forward shock, and the lack of an optical break is attributed to evolving 
microphysical parameters, but with a steady evolution with the shock Lorentz factor. Applying this 
modified forward-shock model to the temporal and spectral properties of several afterglows, we find 
that the absence of the optical break requires a wind-like circumburst medium, which is consistent with 
the origin of long bursts in the death of massive stars. Furthermore, we find and that the evolution 
of microphysical parameters must satisfy equation (\ref{ib}) and that the fraction of the shock energy 
imparted to electrons during the burst is a factor 2--6 larger than at the end of the slow-decay phase. 
This may account in part for the high efficiency of the GRB mechanism.

{\bf Post jet-break phase}.
 For the very few X-ray afterglows where this phase was observed, the decay indices and spectral slopes
are consistent with the expectations from the standard jet model, with or without lateral spreading.
The available optical observations allow the identification of an achromatic break, as expected in the 
jet model, for the afterglows of the Swift GRBs 050525A, 060124, and 060526. However, the optical decay 
during the post jet-break phase of GRB afterglow 050525A is too slow and cannot be accounted naturally 
for by the standard jet model. A similar situation is encountered for the pre-Swift GRB afterglow 010222. 
For these two afterglows, the equality of the pre jet-break optical and X-ray decay indices is consistent 
with the predictions of the standard jet model, where microphysical parameters are constant, but the slow 
post jet-break optical decays require evolving parameters.

\clearpage

\hspace*{-5mm} {\bf REFERENCES} \vspace*{5mm}

 \hspace*{-5mm}
 Bjornsson G. \etal (2002) -- ApJ 579, L59 \\
 Burrows D. \etal (2005) -- preprint (astro-ph/0511039)  \\
 Chevalier R. \& Li Z. (1999) -- ApJ 520, L29 \\
 Curran P. \etal (2006) -- preprint (astro-ph/0610067) \\
 Della Valle M. \etal (2006) -- ApJ 642, L103 \\
 Eichler D. \& Granot J. (2006) -- ApJ 641, L5 \\
 Falcone A. \etal (2006) -- ApJ 641, 1010 \\
 Fan Y. \& Piran T. (2006) -- MNRAS 369, 197 \\
 Fenimore E. \& Ramirez-Ruiz E. (1999) -- preprint (astro-ph/9909299) \\
 Fox D. \etal (2003) -- Nature 422, 284 \\
 Freedman D. \& Waxman E. (2001) -- ApJ 547, 922 \\
 Gendre B. \etal (2006) -- A\&A, submitted (astro-ph/0603431)  \\
 Granot J., Nakar E., Piran T. (2003) -- Nature 426, 138 \\
 Granot J., Konigl A., Piran T. (2006) -- MNRAS, 370, 1946 \\
 Kulkarni S. \etal (1999) -- Nature 398, 389 \\
 Kumar P. (1999) -- ApJ 523, L113 \\
 Kumar P. \& Panaitescu A. (2000) -- ApJ 541, L51 \\
 Lloyd-Ronning N. \& Zhang B. (2004) -- ApJ 613, 477 \\
 M\'esz\'aros P. \& Rees M. (1997) -- ApJ 476, 232 \\
 M\'esz\'aros P., Rees M., Wijers R. (1998) -- ApJ 499, 301 \\
 Nakar E. \& Granot J. (2006) -- MNRAS, submitted (astro-ph/0606011) \\
 Norris J. \etal (1986) -- ApJ 301, 213 \\
 Norris J. \etal (1996) -- ApJ 459, 393 \\
 Nousek J. \etal (2006) -- ApJ 642, 389 \\
 O'Brien P. \etal (2006) -- ApJ 647, 1213 \\
 Paczy\'nski B. \& Rhoads J. (1993) -- ApJ 418, L5 \\
 Paczy\'nski B. (1998) -- ApJ 494, L45 \\
 Panaitescu A., M\'esz\'aros P., Rees M. (1998) -- ApJ 503, 314 \\
 Panaitescu A. \& Kumar P. (2000) -- ApJ 543, 66 \\
 Panaitescu A. \& Kumar P. (2002) -- ApJ 571, 779 \\
 Panaitescu A. \& Kumar P. (2004) -- MNRAS 350, 213 \\
 Panaitescu A. (2005) -- MNRAS 363, 1409 \\
 Panaitescu A. \etal (2006) -- MNRAS 369, 2059 \\
 Ramirez-Ruiz E. \& Fenimore E. (1999) -- A\&AS 138, 521 \\
 Rees M. \& M\'esz\'aros P. (1994) -- ApJ 430, L93 \\
 Rees M. \& M\'esz\'aros P. (1998) -- ApJ 496, L1 \\
 Rhoads J. (1999) -- ApJ 525, 737 \\
 Rossi E., Lazzati D. \& Rees M. (2002) -- MNRAS 332, 945 \\
 Sari R., Piran T., \& Narayan R. (1998) -- ApJ 497, L17 \\
 Spada M., Panaitescu A., M\'esz\'aros P. (2000) -- ApJ 537, 824 \\
 Stanek K. \etal (2006) -- ApJ, submitted (astro-ph/0602495) \\
 Zeh A., Klose S., Kann D. (2006) -- ApJ 637, 889 \\
 Zhang B. \& M\'esz\'aros P. (2002) -- ApJ 571, 876 \\
 Zhang B. \etal (2006) -- ApJ 642, 354 \\

\end{footnotesize}

\end{document}